\documentclass[fleqn]{article}
\usepackage{espcrc2}
\usepackage{graphicx}

\newcommand{\etal}{{\it et al.}}
\newcommand{\etc}{{\it etc.}}
\newcommand{\jc}{$j_c$ }
\newcommand{\REBCO}[4]{{#1}Ba$_{#2}$Cu$_{#3}$O$_{#4}$}
\newcommand{\YBCO}{\REBCO{Y}{2}{3}{7-\delta}}

\newcommand{\Fref}[1]{Fig.\,\ref{#1}}
\newcommand{\Frefs}[1]{Figs.\,\ref{#1}}
\newcommand{\SUST}{Supercond. Sci. Technol. }
\newcommand{\MSE}{Mater. Sci. Eng. }

\title{Growth-related inhomogeneities in bulk melt-grown YBaCuO
crystals}
\author{A.~B.~Surzhenko\thanks{Corresponding author. On leave from
Institute for Magnetism, Kiev, Ukraine. E-mail:
surzhenko@ipht-jena.de}, S.~Schauroth, D.~Litzkendorf,
M.~Zeisberger, T.~Habisreuther, and W.~Gawalek\\ Institut f\"ur
Physikalische Hochtechnologie, Winzerlaer Str. 10, D-07745 Jena,
Germany}

\begin{document}
\begin{abstract}
The homogeneity of a large \YBCO (YBCO) sample fabricated by
top-seeded melt-growth has been studied in details. It was
particularly shown how structural imperfections and
superconducting properties are distributed along the growth front.
These distributions are scarcely explicable within a suggestion
that the a-growth fronts are flat. The obtained data testify to
concave growth fronts.
\end{abstract}
\maketitle

\section{Introduction}
\begin{figure}[!b] \begin{center} \vspace{-20pt}
\includegraphics[width=6.5cm]{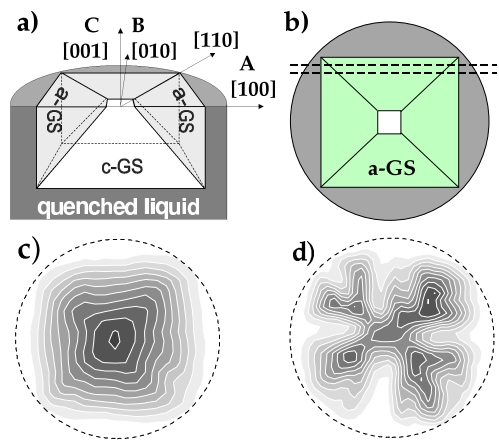}
\vspace{-35pt}
\end{center}
\caption{Schematic diagram (a), (b) for a typical bulk melt-grown
YBCO sample. Compare nearly regular tetragonal shape of isolines
on the remanent flux map obtained after a static FC-magnetization
(c) to their X-like distribution \cite{Surzhenko} remaining after
the magnetic pulse (d).} \label{sample}
\end{figure}

The magnetic flux trapping ability of bulk high-temperature
superconductors (HTS), which promises their wide usage as
cryomagnets \cite{Murakami} operating at liquid nitrogen
temperature $T_{LN}\approx 77\,K$, depends on their critical
current density \jc and the length scale $d$ over which it flows
\cite{Bean}. Therefore, rather large sizes ($d\gg 1\,cm$) are
required for existing \REBCO{RE}{2}{3}{7-\delta} (RE: Y, Sm, Eu,
Nd, \etc) crystals with \jc$(T_{LN})\approx 10^4\,A/cm^2$  to trap
the peak fields $\mu_0H\geq 1\,T$ and, consequently, to seem more
attractive than conventional permanent magnets. However, so large
distances, which the growth front has to cover from the seeding
point, inevitably result in negative changes of growth conditions.
That is why Dewhurst \etal\ observed a certain degradation of the
crystal microstructure in the radial direction (from the crystal
center towards the edges) and, because of a short coherence length
$\xi$ (e.g., $\xi_{c}\approx 0.3\,nm$ and $\xi_{ab}\approx 2\,nm$
in the \YBCO\ (YBCO) material), that of superconducting properties
\cite{Dewhurst}. Apart from the radial inhomogeneities the lateral
ones were also recently revealed \cite{Surzhenko}. In particular,
the lateral distribution of a flux  trapped by the YBCO crystal
after its magnetization by short magnetic pulses, was demonstrated
to ensue from its growth in five growth sectors (GS), i.e. the
sectors grown on different habit planes (see \Fref{sample}).

This paper reports the detailed study how superconducting
properties of a large YBCO sample vary along the a-habits, (100)
and/or (010), and correlates these  effects with the spatial
distribution of the structure inhomogeneities, i.e. grain
boundaries (GB), Y$_2$BaCuO$_{5}$ ('211') inclusions and oxygen
deficient zones. The obtained data are hard to explain in terms of
flat habitus planes, these testify to concave growth fronts.

\section{Experimental results}
The YBCO sample with a diameter of $30\,mm$ and a thickness of
$20\,mm$ was fabricated from the mixture of commercially purchased
powders (Solvay GmbH, Germany) with 1\,wt.\% CeO$_2$ by usual
peritectic solidification, which details are given in
Refs.\,\cite{Surzhenko,Litzkendorf}, and post-annealed in oxygen
atmosphere. The YBCO crystal appeared a large single grain
(please, note nearly regular tetragonal shape of isolines on the
remanent flux map, \Fref{sample}c) which grew as shown in
\Frefs{sample}(a),(b), i.e. in four a-GSs grown on the (100),
(010), ($\bar{1}$00) and (0$\bar{1}$0) habit planes and c-GS which
habit, (001), is perpendicular to the cylinder axis \cite{Diko}.

\begin{figure}[!t] \begin{center}
\includegraphics[width=6cm]{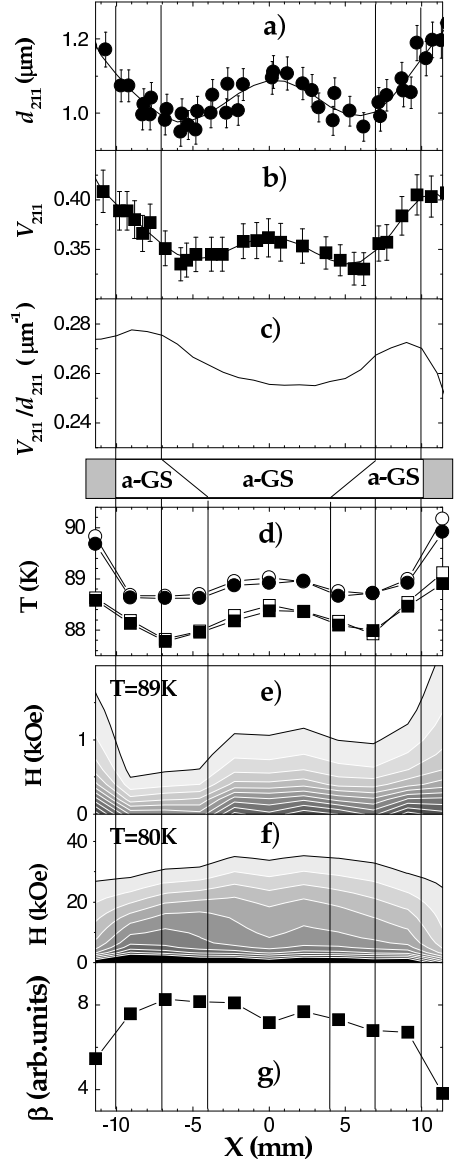} \vspace{-35pt}
\end{center}
\caption{Variation of some structural (a)--(c) and superconducting
(d)--(g) properties along the YBCO slice (see details in the
text).} \vspace{-25pt} \label{slice}
\end{figure}

A slice of width $2.8\,mm$ (marked in \Fref{sample}(b) by dashed
lines) was first cut through a whole sample and a structure of the
YBCO material nearby its top surface (i.e. inside the a-GSs) was
studied in a polarized light. We found out that the boundaries
(a-a-GSBs) of two adjacent a-GSs  are free of subgrains, the
subgrain-free band of $\approx 1\,mm$ thickness bears from the top
surface down to the c-GS. The more the distance from a-a-GSBs, the
larger becomes the polarized light contrast which follows an
increase of mis-orientation angle between neighboring subgrains
\cite{Diko}. The spatial distribution of 211-inclusions along the
a-habit presented in \Frefs{slice}(a), (b) seems also closely
related to the growth of the YBCO crystal in GSs. Both their
average size $d_{211}$ and volume fraction $V_{211}$ dependencies
have nicely symmetric, W-like shape.

Similar behavior is registered for the transition temperature
$T_c$ presented in \Fref{slice}(d). To obtain these data as well
as distributions of critical current densities $j_c(H\parallel c)$
at different temperatures shown in \Frefs{slice}(e), (f), the
slice was diced into 11 samples (each of sizes $2.8 \times 2
\times 2.4\,mm$), which magnetic properties were measured with
ac-susceptometer (0.1\,Oe, 20\,KHz) and vibrating sample
magnetometer (Oxford Instruments, VSM\,3001). In fact, the plot
(d) presents two temperatures where the transition starts ($T_c$)
and ends. These are determined as the points wherein
ac-susceptibility $\chi$ (normalized to unity) approaches the
values $-0.1$ (circles) and $-0.9$ (squares), respectively. The
open points are obtained when $H\parallel c$, the closed ones
correspond to $H\parallel ab$.

The lowest-level $j_c$ isoline at $T=89\,K\rightarrow T_c$ (solid
black curve in \Fref{slice}e), i.e. the line connecting the
irreversibility points $H_{IL}$ of studied samples, confirms the
W-like distribution of $T_c$ which, according to a common opinion
\cite{Sandiumenge}, reflects a difference in an oxygen content.
Since an oxygen is easy to penetrate into imperfect material
containing cracks and subgrain boundaries, a quenched liquid
exhibits the higher $T_c$. In more structurally perfect crystal
$T_c$ decreases anywhere except for the central part ($X\approx
0$) of the a-GSs. Thus, this area may be expected somewhat alike
in structure to a quenched liquid.

At lower temperatures an oxygen influence on superconducting
properties is no longer dominant (see \Fref{slice}f). The
$j_c\,vs\,H$ dependencies look more homogeneous (e.g, within the
YBCO crystal $j_c(H=0,77\,K)=55\pm 5\,KA/cm^2$) and are well
fitted by the '$-1/2$'-power law ($j_c\sim \beta H^{-1/2}$,  where
$\beta\sim V_{211}/d_{211}$), which is attributed to the
211-inclusions \cite{Sandiumenge}. Besides, the spatial
distribution of $\beta$ (\Fref{slice}g) along the a-growth front
looks quite similar to the value $V_{211}/d_{211}$ which is
presented in \Fref{slice}(c) and directly calculated from the
$V_{211}$ and $d_{211}$ polynomial fit (solid lines). Some
diversities are easy to explain: in contrast to $V_{211}/d_{211}$,
$\beta\,vs\,X$ curve corresponds to the 211-distribution
\emph{averaged} over the finite slice thickness, $2.8\,mm$.

\begin{figure}[!tbp] \begin{center}
\includegraphics[width=6cm]{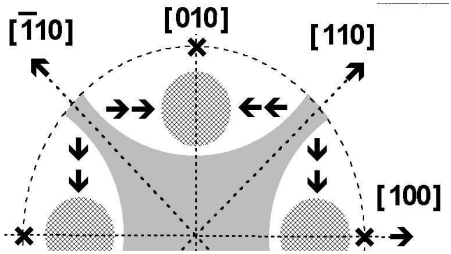} \vspace{-30pt}
\end{center}
\caption{A feasible scenario of the YBCO crystal growth. Gray
color denotes the YBCO material, shaded regions correspond to the
clouds wherein 211-inclusions are pushed by secondary growth
fronts (marked by double arrows).} \vspace{-25pt} \label{concave}
\end{figure}

\section{Conclusions}
In summary, one can conclude that structural and superconducting
properties agree well each with other, but both are hard to
explain in terms of a suggestion that a-habits are flat.  At the
same time, these features are readily derived from the hypothesis
of concave growth fronts (see \Fref{concave}). Within this
approach, the YBCO crystal {\it first} grows along the X-like
'backbone', i.e. in the [110], [1$\bar{1}$0], [$\bar{1}$10] and
[$\bar{1}\bar{1}$0] directions (dashed arrows), whereas the {\it
secondary} growth fronts, which respond for material inside a-GSs,
are parallel to the [100] and [010] directions (shown by small
double arrows). Since during the growth process the crystal
scarcely reveals so strong pillow distortion as this shown in
\Fref{concave}, the secondary growth fronts are expected to move
much \emph{faster} than those forming the 'backbone'. It certainly
results in large mechanical stresses inside the a-GSs and,
therefore, an appearance of subgrains trying to reduce these
stresses. In terms of the pushing/trapping theory \cite{Kim95},
faster motion of secondary growth fronts may also explain an
increased values of $d_{211}$ and $V_{211}$ shown in
\Frefs{slice}(a), (b). At last, the square shape of YBCO crystals,
which is well confirmed by, for instance, the remanent flux map in
\Fref{sample}(c), means that growth process stops when the
'backbone' reaches the size of a crucible. If the growth mechanism
is different from the mentioned above, the crystal could grow
until \emph{a whole} crystal fits in a crucible, i.e. in a circle.

\section*{Acknowledgements}
The work was supported by the German BMBF under the project
No\,13N6854A3. The authors would like to thank Z.\,H.\,He and
P.\,Diko for illuminating discussions as well as R.\,M\"uller for
his continual encouragement throughout the present work. We are
also indebted to P.\,Dittman, M.\,Arnz and Ch.\,Schmidt for
technical support.

\end{document}